**Title**

Quantum Phases and Spin Liquid Properties of 1T-TaS$_2$

(Short title: Spin Liquid Properties of 1T-TaS$_2$)

**Authors**

Samuel Mañas-Valero[1], Benjamin M. Huddart[2], Tom Lancaster[2], Eugenio Coronado[1], Francis L. Pratt[3]*

**Affiliations**

[1] Universidad de Valencia (ICMol), Catedrático José Beltrán Martínez, 46980 Paterna, Spain.

[2] Centre for Materials Physics, Durham University, Durham DH1 3LE, United Kingdom.

[3] ISIS Neutron and Muon Source, STFC Rutherford Appleton Laboratory, Didcot OX11 0QX, United Kingdom.

* Correspondence to: francis.pratt@stfc.ac.uk

**Abstract**

Quantum materials exhibiting magnetic frustration are connected to diverse phenomena including high-$T_c$ superconductivity, topological order and quantum spin liquids (QSLs). A QSL is a quantum phase (QP) related to a quantum-entangled fluid-like state of matter. Previous experiments on QSL candidate materials are usually interpreted in terms of a single QP, although theories indicate that many distinct QPs are closely competing in typical frustrated spin models. Here we report on combined temperature-dependent muon spin relaxation and specific heat measurements for the triangular-lattice QSL candidate material 1T-TaS$_2$ that provide evidence for competing QPs. The measured properties are assigned to arrays of individual QSL layers within the layered charge density wave structure of 1T-TaS$_2$ and their characteristic parameters can be interpreted as those of distinct $Z_2$ QSL phases. The present results reveal that a QSL description can extend beyond the lowest temperatures, offering a new perspective in the search for novel quantum materials.

**Teaser**

The commensurate charge-density-wave in 1T-TaS$_2$ gives a spin ½ triangular lattice that supports quantum spin liquid states.

# MAIN TEXT

**Introduction**

The idea of destabilizing magnetic order by quantum fluctuating resonating-valence-bonds (RVB) originated with Anderson in 1973 (*1*). Systems with frustrated interactions are particularly susceptible to these effects and much work on highly frustrated systems and their quantum spin liquid (QSL) phases has followed. In parallel with extensive theoretical work (*2–4*), many candidate QSL materials have been proposed. For triangular lattices these include molecular (*5–8*) and inorganic (*9–11*) compounds. However, to date, little experimental effort has been devoted to testing for crossover between different topological phases in a QSL, or to exploring the nature of the fractionalized spinon excitations (bosons, fermions or anyons) (*12*). This report concerns 1T-$TaS_2$, which is a layered compound with a series of charge-density-wave (CDW) instabilities terminating in a fully commensurate CDW phase (C-CDW) below 200 K. In the C-CDW state the distortion pattern forms a triangular-lattice of Star-of-David (SD) clusters, each cluster containing 13 Ta atoms and one unpaired spin-½ (Fig.1A). It shows no evidence for magnetic ordering and was one of the triangular-lattice materials that originally inspired the RVB theory (*1*).

The electronic and magnetic scenario offered by 1T-$TaS_2$ is far from being simple and is currently being revisited. Theoretical studies have shown that a single layer of 1T-$TaS_2$ is indeed a QSL (*13*) and experimental results in the selenium counterpart, 1T-$TaSe_2$, have revealed that single layers are Mott insulators that behave as QSLs (*14, 15*). However, considering bulk crystals, where the layers are stacked in the out-of-plane direction, new challenges for the interpretation of its properties appear. Taking into account band theory, 1T-$TaS_2$ should be a metal, since there are an odd number of electrons per unit cell. As it has been experimentally observed to have an insulating state below 200 K, the Mott mechanism has been invoked to explain this apparent contradiction and, thus, the possibility of a QSL has emerged (*13*). However, if the layers form dimers in the out-of-plane direction, the unit cell doubles and there are an even number of electrons, thus transforming the system to a conventional band insulator where the QSL scenario has no place (*16, 17*). Several recent studies have made further exploration of the evidence that 1T-$TaS_2$ is a QSL, both from experimental and theoretical perspectives (*13, 18–21*), but some alternative scenarios have also been proposed, such as a Peierls mechanism or the formation of domain wall networks, among others (*22–24*).

Although the QSL may seem directly linked to the presence or absence of dimerization of the layers in the out-of-plane direction, there is still another possibility, since the out-of-plane stacking order can be arranged in different configurations and exhibit a temperature dependence, as recently proposed (*25, 26*). Thus, below 200 K the layers can start off being uncoupled, allowing a QSL to exist, but then they can become coupled progressively while cooling down, moving towards a band-insulator at low temperatures if all the layers become dimerized. However at intermediate temperatures, there is a scenario where observed QSL properties may be the result of coexisting dimerized and undimerized layers. This relatively unexplored picture is examined more closely in the present study, where we report on muon spin relaxation (µSR) and specific heat as a function of the temperature on high quality samples of 1T-$TaS_2$. We observe distinct regions versus temperature for both of these properties. We discuss these different observed thermal regions in relation to possible changes in the out-of-plane stacking order of the layers and also in relation to some models of closely competing QSL states that are currently found in the literature.

# Results

*Muon spin relaxation (μSR)*

μSR provides a good way to check a candidate QSL for magnetic ordering and several previous reports using zero-field μSR (ZF-μSR) have confirmed the absence of magnetic ordering in 1T-TaS$_2$ down to 20 mK (*18–20*). The muon probe is fully spin polarized on implantation and the forward/backward asymmetry of the detected muon decay positrons, $a(t)$, reflects the time dependent polarization of the muon spin. In the ZF measurements (Fig. 2A), the relaxing component of $a(t)$ was fitted to the product of a Gaussian and a Lorentzian term. The Gaussian term reflects nuclear dipolar relaxation contributions and was estimated at high temperature and then kept fixed as the temperature varied. The relaxation rate of the Lorentzian term λ reflects the electronic contribution to the muon spin relaxation that varies significantly with temperature. The electronic contribution is in a fast fluctuation regime where λ is proportional to the electronic spin fluctuation time τ (*27*). μSR can also measure critical parameters (*6*) that can be compared against QSL models. A QSL can support fractionalized spinon excitations and various experimental properties will depend on the characteristics of these spinons. Here we use μSR in longitudinal magnetic fields (LF-μSR) to provide information on the dynamics of spinons diffusing through the triangular lattice (Fig.1A). The coupling of the muon to the spinons is determined by the stopping sites (Fig.1B) and associated hyperfine interactions, which we have computed using density functional theory (DFT).

The muon spin relaxation is relatively slow in this material (Fig.2A), as found previously (*18–20*). A detailed study of λ($B_{LF}$) at fixed $T$ provides information about the spin fluctuations (*28*). This method has been used in μSR for studying the propagation of spinons in spin-½ Heisenberg antiferromagnetic chain systems (*29*), where 1D diffusive motion leads to a weak power law dependence for λ($B_{LF}$), and the method can be extended to systems of higher dimensionality (*28*). Figure 2B shows example data at 120 K, fitted against spinon diffusion models on 1D, 2D or 3D lattices. The 2D model clearly provides the best fit. Figure 3A shows the 2D diffusion rate $D_{2D}$ obtained from fits to the LF-μSR scans. Several distinct thermal regions are observed and effective thermal power laws are used to describe the data in the different regions. The power law parameterizations will be useful for making comparison with QSL models, which make their predictions in terms of power laws. On cooling through $T_{nC}$ and entering the C-CDW phase, $D_{2D}$ rises significantly, consistent with the presence of a QSL state that supports diffusing spinons. This higher temperature region of the C-CDW phase is labelled III and on cooling, $D_{2D}$ continues rising until a maximum occurs at $T_0 \sim 110$ K. The region below $T_0$ labelled II shows a relatively strong power law $D_{2D} \propto T^{n_D}$ with $n_D = 1.74(14)$. Below $T_1 \sim 25$ K another region, labelled I, is found, in which $n_D = 0.47(17)$. Fig.3B shows the diffusive signal amplitude parameter $f_0$, which is the product of $f_{diff}$, the fraction of spectral weight in the diffusion process, and $f_{para}$, the fraction of muon sites coupled to unpaired electronic spins. The values for $f_0$ have been obtained using the muon hyperfine coupling calculated from DFT analysis $\bar{A} = 4.5(6)$ MHz (see supplement). It can be seen that the $f_0$ parameter also shows multi-region behavior, with region boundaries matching those of $D_{2D}$.

*Specific Heat*

An independent measurement of the behavior of the QP regions I, II and III is given by the power law electronic contribution to the specific heat $C_p^{el} \propto T^{n_C}$. Details of the measurements and their analysis are provided in the supplement. These provide an experimental value of $n_C$ for each region (Fig.4). The overall behavior of the specific heat is summarized in Fig.4A. The border between regions II and III is clearly signified by a maximum value of $C_p/T$ at $T_0$, corresponding to the distinct peak found in the μSR parameters of Fig.3. In contrast to the strong feature at $T_0$, the border between regions I and II at $T_1$ is not so obvious because the lattice term is rapidly varying in this region. In the low $T$ region the lattice contribution is taken to follow the standard asymptotic $T^3$ Debye form and the power law $n_C$ for the electronic contribution in region I can then be obtained from a fit, as shown in the associated scaling plot of Fig.4B. The obtained value of 1.46 is significantly larger than expected for a spinon Fermi liquid, where the predicted value is 1, reducing to 2/3 when the system is close to a quantum critical point (QCP) (*30*). For regions II and III the lattice contribution is modelled by the sum of an anisotropic Debye term and two Einstein terms, as detailed in the supplement. The peak in $C/T$ is then seen to be the result of $n_C$ reducing by around 1 on warming through $T_0$, as shown in Fig.4C.

**Discussion**

We consider here possible interpretations of the different regions observed in these measurements. First we look at how the experimentally obtained power laws compare with the critical parameters of appropriate QSL models. Next we consider the nature of the transitions between the regions and a possible interpretation in terms of a delicate balance between competing QSL phases, as found in numerical studies of the triangular lattice Heisenberg antiferromagnet. Finally we consider the role of the degree of dimerization of the layers defined by the interlayer stacking of the SD clusters.

*Critical parameters and QSL models*

The many possible 2D QSL states for a triangular lattice can be broadly classified (*13*, *31*) as $SU(2)$, $U(1)$ or $Z_2$. Of these, the $Z_2$ states are expected to be most stable (*4*, *13*) and there are well-developed theories for describing the quantum critical (QC) properties of $Z_2$ QSL phases (*32–34*). Thus $Z_2$ models provide a natural starting point for interpreting the data. A key feature of QC phases is that they occupy extended regions of $T$, in contrast to classical critical regions, that generally require $T$ to be very close to the transition point. For a $Z_2$ QSL the spinon transport in the QC regime (*34*) is determined by a momentum relaxation time that follows power law in $T$ of 2/ν-3, where ν is the correlation length exponent. Applying the Einstein diffusion relation, the power law for the diffusion rate is then larger by one

$$n_D = 2/\nu - 2. \tag{1}$$

Besides this critical power law, there may also be present an additional non-critical contribution, related to an energy dependence for the spinon density of states (DOS). This can be taken into account by including an additional parameter $q$

$$n_D = 2/\nu - 2 + q. \tag{2}$$

The value $q = 1$ describes a QSL with a linear DOS, such as would be found for a nodal excitation spectrum, whereas $q = 0$ for a QSL with constant DOS. The ν exponent in the QC model (*32–34*) is that of the O(4) criticality class, calculated to be 0.748(1) (*35*).

For the specific heat, the corresponding critical exponent α can be obtained from ν via the hyperscaling relation $\alpha = 2 - 3\nu$, giving $\alpha = -0.244(3)$. The power law for the specific heat is then taken to follow the form

$$n_C = 1 - \alpha + q \qquad (3)$$

Experimental values for α and ν are compared with the model values in Fig.5. For region I the experimental values for α and ν are both reasonably consistent with the $Z_2$ QSL model with $q = 0$. For region II, both exponents closely matches theory if $q = 1$, suggesting a $Z_2$-linear QSL. In region III it is found that $q = 0$ again gives the closest match to the theoretical exponent. We note that the strong negative power found for $D_{2D}$ in region III (Fig.3A) is not predicted within this model. We have considered whether muon diffusion might be responsible for this behavior, but the energy barriers between adjacent sites in the *ab* planes and also from one side of the layer to the other are found to be too large to support diffusion in this temperature range (see supplement). It is more likely that the fall reflects inter-plane spin correlations sensed by the muon (see supplement), as well as the influence of charge excitations, which would be expected to become significant in this region that borders on the nC-CDW phase. We note that a negative power might be the result of a $U(1)$-gapless QSL state with fermionic spinons (*36*), but the predicted exponent $n_D = -1/3$ is very much smaller than measured here (Fig.3A) and the corresponding specific heat prediction (*30*) of $\alpha = 1/3$ is not found (Fig.5B).

*The nature of the phases and their transitions*

One question concerns the nature of the transitions between the regions, in particular whether they are crossovers or phase transitions. The absence of any sharp features in the specific heat in the whole C-CDW/QSL region rules out first-order phase transitions. A conceptual framework that could be used for interpreting the transitions in terms of underlying quantum phase transitions (QPTs) (*37, 38*) is illustrated in Fig. 6. All three phases are taken to lie within the QC regime associated with a QCP separating QSL and AF ground states, governed by control parameter $g_1$ (Fig.6A). A possible association of $g_1$ could be made with ring-exchange, which is known to destabilize the magnetic phase of triangular-lattice Mott-Hubbard insulators (*39*). The energies of the competing QSL states are assumed to depend on a second control parameter $g_2$ and at $T = 0$ continuous second-order QPTs would be expected between pairs of QSL states (Fig. 6B). Moving to the finite $T$ case, appropriate to our measurements, several new factors come into play. Firstly, since the excitation spectrum is different between the adjacent QSL phases, as indicated by their difference in $q$, there will in general be a difference in entropy between the phases at the transition temperature, which would tend to turn the continuous second order transition into a discontinuous first-order transition. A second factor is that at finite $T$ we expect the QPTs to evolve into broad QC regions separating the QSL phases, with the transitions becoming crossovers. In the current system it appears that the second mechanism dominates over the first. In this scenario, $g_2$ is taken to have a $T$ dependence that produces smooth crossover transitions between the QSL states. The finite temperature transitions at $T_0$ and $T_1$ then would directly correspond to multicritical QCPs in a suggested underlying $(g_1, g_2)$ quantum phase diagram for $T = 0$, as shown schematically in Fig.6B.

*Comparison with a triangular-lattice QSL phase diagram*

Spinon pairing instabilities leading to $Z_2$ spin liquid states have been discussed previously in several studies (*40–43*). In general, there are 64 possible symmetric $Z_2$ spin liquid states for the triangular lattice (*44–46*). Some of the more stable states have been revealed via variational studies and it is useful to discuss the multiple phases revealed by our measurements in relation to previously reported phase diagrams from these studies. The most comprehensive study to date is by Mishmash *et al.* (*40*). This uses a model Hamiltonian which includes terms representing the nearest-neighbor exchange *J*, the next-nearest-neighbor exchange $J_2$ and the ring-exchange *K*. The parameters $J_2$ and *K* are expressed in units of *J*. There is general agreement from different studies that for this type of model the AF state gives way to the $U(1)$ spin liquid state with a spinon Fermi surface when *K* exceeds values of order 0.2-0.25 (*39, 40, 43, 47*). At lower *K* a wider range of states have been suggested. Figure 6C shows a representation of the part of the phase diagram obtained by Mishmash *et al.* (*40*) that is most relevant to a Mott-Hubbard insulator, where the exchange parameters are determined by the *U/t* ratio. In this study a BCS-like *d*-wave paired $Z_2$ spin liquid was found over an extended region of the phase diagram (*40*) (Fig. 6C). The properties of the nodal state were discussed by Grover *et al* (*43*). Below a characteristic pairing temperature it shows a linear ($q = 1$) DOS. In their study Mishmash *et al* (*40*) considered six possible pairing modes and found a second paired state, labelled *d + id*, that is very close in energy to the nodal state and the most stable state for a small region of parameter space indicated in Fig. 6C. Having such strongly competing states corresponds well with the picture put forward in Fig. 6B and a clear mapping of the phenomenological $g_1$ and $g_2$ control parameters to the $K$-$J_2$ space is suggested by the structure of the phase diagram (Fig.6C, dashed lines). The *d + id* state has quadratic dispersion around the central Γ point (*40*) which gives a constant DOS state with $q = 0$. A possible cooling path and assignment of the two lowest temperature thermal phases is indicated by the purple arrowed line in Fig. 6C. Within this picture phase II could be assigned to the $q = 1$ nodal *d*-wave state and phase I could be assigned to the $q = 0$ quadratic *d + id* state.

The scenario of Fig.6C would place *K* in the region of 0.12, corresponding to a *U/t* ratio of order 11. This can be compared to the *U/t* estimates of 7 to 8 for the well-studied triangular-lattice QSL system κ-(ET)$_2$Cu$_2$(CN)$_3$, with corresponding *K* values in the region 0.5 to 0.3. Although model phase diagrams such as Fig. 6C provide a good starting point for discussion, further distinct properties of 1T-TaS$_2$ need also to be taken into account for a more complete picture, as discussed in the next section.

*Interlayer stacking*

The mode of interlayer stacking of the SD clusters has an important bearing on the local electronic properties and, in particular, the emergence of insulating behavior. However, this aspect of 1T-TaS$_2$ remains poorly understood. Recent experiments address possible origins for the insulating state, from the conventional Mott scenario (*19*) to layer dimerization (*48*), unit-cell doubling (*24*) or a Peierls transition (*22*). Synchrotron X-ray studies reveal significant disorder in the interlayer stacking of the SD clusters at 100 K from the presence of diffuse scattering (*48*), suggesting a transition from a band insulator at low temperatures to a Mott insulator at higher ones (*25*). Besides a broad diffuse background, diffuse peaks were observed at positions suggesting correlations in the stacking sequence on length scales of three and five layers. In contrast, data taken for the nC-CDW phase at 300 K show a simple three-fold periodicity in the interlayer stacking (*48*). A scanning tunneling microscopy study (*24*) found two types of cleaved surface at 77 K: large-gap surfaces that were assigned to spin-paired bilayers and narrow-gap surfaces assigned to unpaired Mott-gapped layers, that could support the QSL state. The large-gap surfaces were assigned to surface bilayers (A-mode stacking with SDs aligned between layers). The weaker-gap surfaces were assigned to a surface layer stacking denoted as a C-mode, for which the central Ta site of the SD of the lower layer is located below the Ta site in one of the tips of the SD of the upper layer. A regular stacking sequence of the form ACACAC is one possibility (Fig. 7A). If the electronic stacking along the c axis is more complex, e.g. taking the form ACACCAC (or, in general, …ACA⊥⊥AC…, where ⊥ is a layer with a stacking different than A), then a fraction of the cleavage planes will be unpaired monolayers. The experimental ratio of three to one found in ref. (*24*) would then be consistent with an average density of electronic stacking defects around this one in seven level (Fig. 7B). For muons implanting in such a sequence, two out of seven interlayer sites will see the unpaired layer and couple to the paramagnetic electronic relaxation of the QSL layers and the remaining five sites will see the diamagnetic spin-paired bilayers, showing only nuclear relaxation, thus $f_\text{para} = 2/7$ in this case. Figs. 7C and 7D show examples of local configurations with increasing QSL layer concentrations and correspondingly larger values for $f_\text{para}$.

*Comparing alternative scenarios for the higher temperature regions*

The $f_0$ parameter shown in Fig. 3B can provide valuable information about evolution of the electronic state with $T$. In region I it is essentially constant, indicating a stable electronic state and allowing the QC model to be applied with confidence. On the other hand, the significant changes of $f_0$ seen in regions II and III provide a greater challenge for the overall interpretation. The largest value of $f_0$ is found at $T_0$, where it approaches its maximum possible value of 1, indicating that both $f_\text{para}$ and $f_\text{diff}$ approach 1 here. The reason for the drop in $f_0$ in region III above $T_0$ is not entirely clear, but besides reflecting AF interlayer correlations at the muon site, it may also reflect the increasing influence of charge excitations, which are not included in the basic spin-only model. The origin of the drop in $f_0$ on the lower side of $T_0$ when cooling through region II is also a key question. In one scenario we could assume that $f_\text{diff}$ is 1 at all $T$, so that the drop in $f_0$ is entirely due to a reduction in $f_\text{para}$ at low $T$. This could be consistent with the STM study (*24*), as discussed in the previous section. In this scenario one could propose that the $q = 1$ value obtained for the specific heat in region II would not reflect a QSL state with a linear DOS, but rather a linear increase in the concentration of the QSL layers above $T_1$, reaching a maximum at $T_0$. This scenario would not however be consistent with the observed large change in $n_D$ at $T_1$, since the obtained 2D diffusion rate is a property of an individual QSL layer and it would not be expected to directly depend on $f_\text{para}$. The opposite scenario for the drop in $f_0$ is that $f_\text{para}$ remains at or close to 1, so that the changes in $f_0$ observed in region II are primarily due to changes in $f_\text{diff}$. In this case the $q = 1$ value in region II would reflect an underlying QSL having a linear DOS, as discussed earlier. This latter scenario would then be consistent with both the diffusion rate and specific heat data obtained in the present study.

*Conclusion*

The main result of our study is that the combined local-probe LF-µSR and bulk specific heat measurements in 1T-TaS$_2$ within the C-CDW regime support the existence of QSL layers within the layered structure. The overall thermal behavior is however found to be complex, with three distinct thermal regions observed in the data. For region I, the most stable region that occurs at low temperature, it is found that both sets of measurements are consistent with the QC regime of a $Z_2$ QSL model. The specific heat in region III is also consistent with this model. In contrast, region II shows an additional *T*-linear contribution to the measured properties, which most likely reflects the presence of a competing QSL state with a nodal excitation spectrum. Such competing nodal phases have previously been found in various theoretical studies of the S=1/2 triangular lattice Heisenberg antiferromagnet. Compared to the usual triangular lattice QSL systems with fixed lattice structure for the spin sites, the 1T-TaS$_2$ system has a unique additional degree of freedom related to weak interlayer correlations affecting the mode of stacking between the SD layers. Any tendency within the stacking to form dimerized bilayers will lead to a non-magnetic contribution from these bilayers. From our data we conclude that in our sample a large fraction of the layers are undimerized at the boundary between regions II and III and still only a minor fraction become dimerized at low temperature.

**+Materials and Methods**

1T-TaS$_2$ crystals were grown by conventional chemical vapour transport and characterized using elemental analysis, ICP mass spectrometry and powder X-Ray diffraction. Heat capacity and transport measurements were performed using a Quantum Design PPMS. The µSR experiments were carried out on the HiFi instrument at the ISIS Neutron and Muon Source. Detailed measurements of the field and temperature dependence of the muon spin relaxation were made in this study, counting $5 \times 10^7$ muon decay events in each µSR spectrum to get sufficient accuracy in the fitted parameters to estimate critical exponents that can be compared with those of QSL models. Results were analysed using the WIMDA program (*49*). The error bars in Fig.2A represent the standard error derived from the counting statistics. The error bars in Fig.2B and Figs.3-5 represent the standard errors of the fitted parameters. Further details regarding the sample preparation and characterization, magnetic susceptibility, specific heat and transport measurements, muon measurements, DFT details and general considerations are given in the supplementary information.

**Acknowledgments**

**General**: Computing resources were provided by STFC Scientific Computing Department's SCARF cluster. The authors thank Marco Evangelisti for assistance and helpful discussions with specific heat measurements.

**Funding:** This work has been supported by the European Commission (COST Action MOLSPIN CA15128 and ERC AdG Mol-2D 788222 to E.C.). F.P and T.L. acknowledge the support of UK EPSRC (EP/N024486/1 and EP/N024028/1). B.H. thanks STFC for support via a studentship. E.C. and S.M.-V. acknowledge the Spanish MINECO (Project MAT2017-89993-R co-financed by FEDER and the Unit of Excellence "Maria de Maeztu" MDM-2015-0538) and the Generalitat Valenciana (Prometeo Programme). S.M.-V. thanks the Spanish Government for the doctoral scholarship F.P.U (FPU14/04407).

**Author contributions:** SMV prepared and characterized the samples under the supervision of EC. The muon measurements were carried out by FLP, SMV and BMH. Magnetic susceptibility, transport measurements and specific heat was measured by SMV. FLP and SMV analyzed the data. BMH, FLP and TL performed the muon site analysis. FLP devised the interpretive framework and drafted the initial manuscript. All authors reviewed and revised the manuscript.

**Competing interests:** The authors declare no competing interests.

**Data and materials availability:** The muon raw data files are available at DOI:10.5286/ISIS/E/RB1810583. Other parts of the data are available from the corresponding author upon reasonable request.

# Figures and Tables

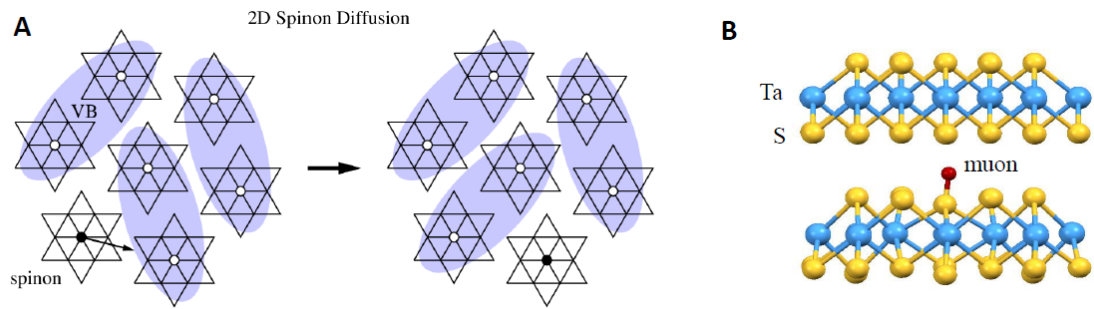

**Fig. 1. Excitations in 1*T*-TaS$_2$ and the Muon Probe Site.** (**A**) The basic hopping step for 2D spinon diffusion in a QSL involves rearrangement of the valence bonds (VB). (**B**) A typical interstitial muon probe site obtained using DFT.

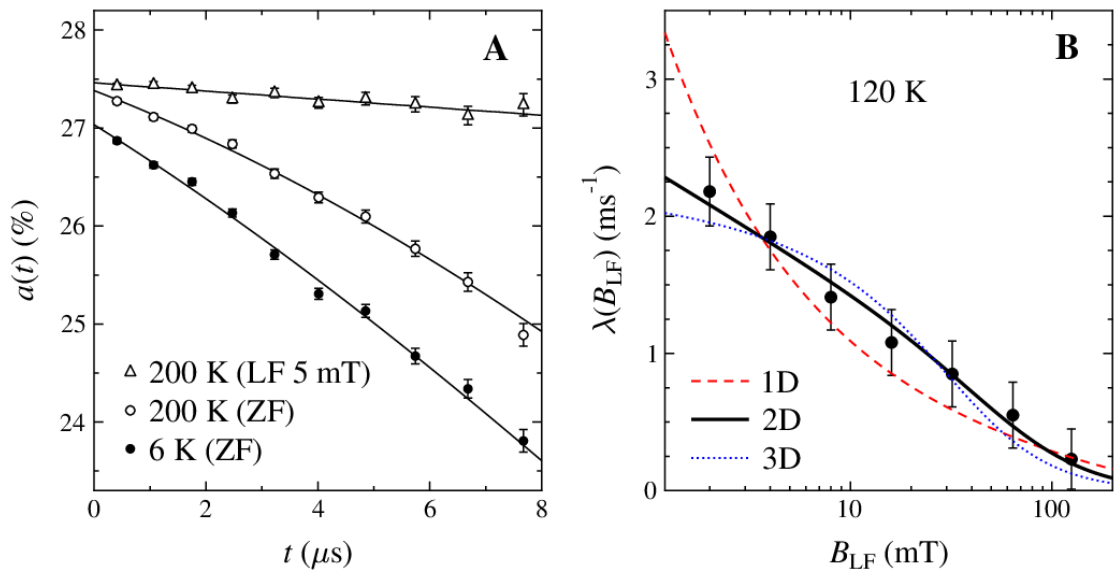

**Fig. 2. Muon Spin Relaxation.** (**A**) ZF-µSR at 200 K and 6 K and the quenching effect of weak LF on the relaxation rate at 200 K. (**B**) LF dependence of the relaxation rate at 120 K with fits to 1D, 2D and 3D spinon diffusion models.

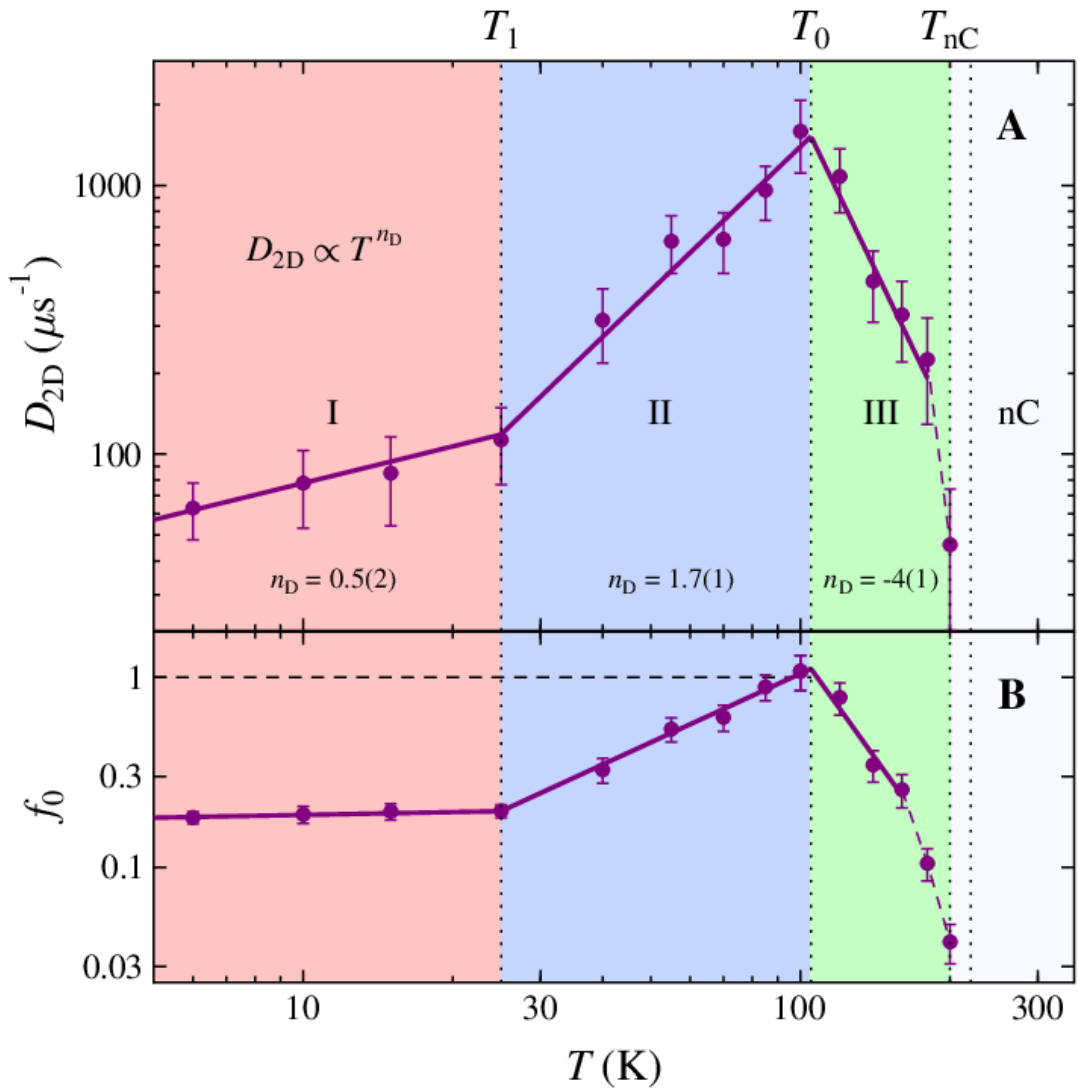

**Fig. 3. Spinon diffusion obtained from LF-μSR in the C-CDW region below $T_{nC}$.** (A) Temperature dependence of the 2D spinon diffusion rate $D_{2D}$. (B) The scaling parameter for the amplitude of the diffusive signal (taking $\bar{A} = 4.5$ MHz). Three distinct regions are found in the measurements, labelled I, II and III.

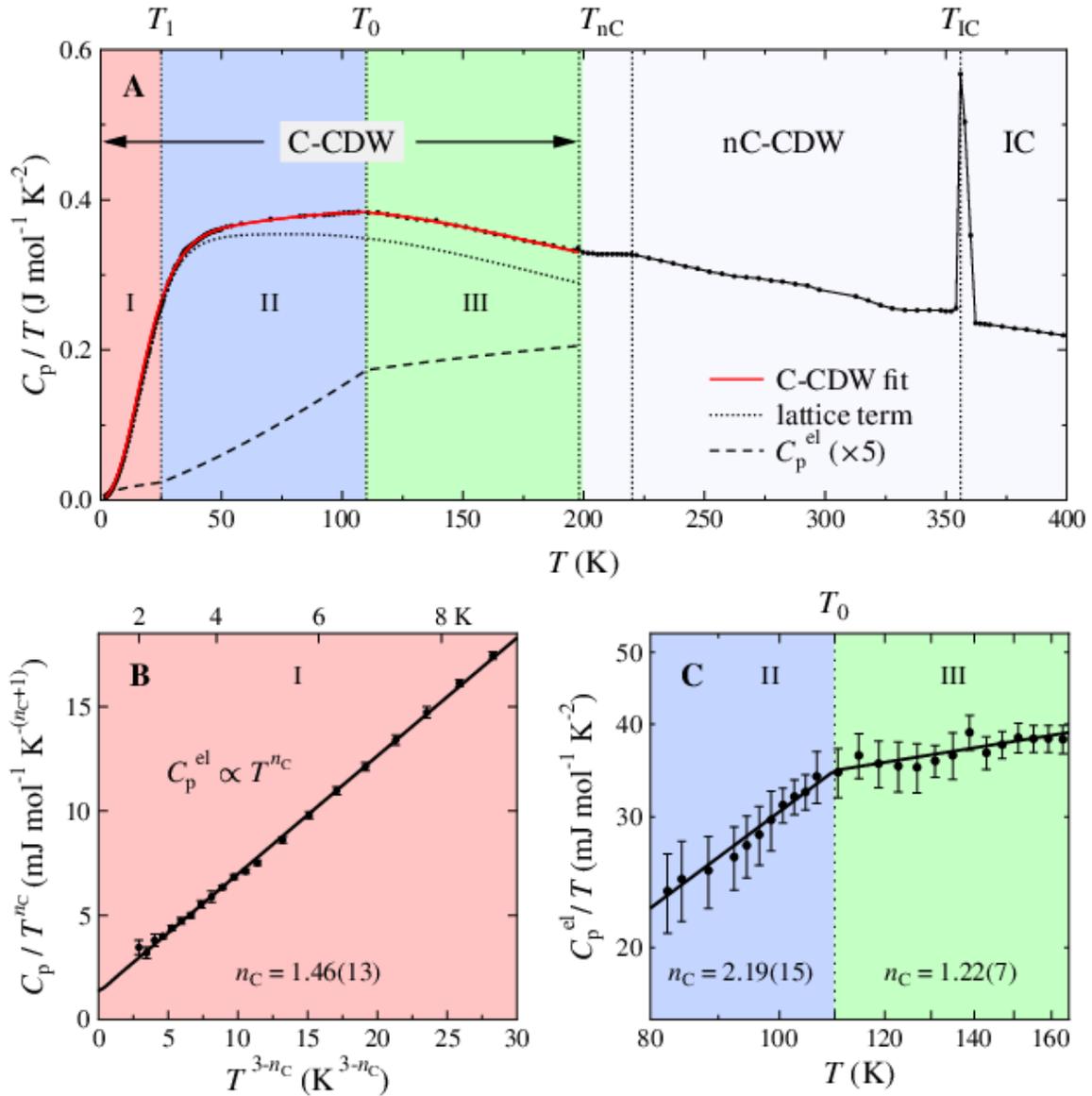

**Fig. 4. Specific Heat.** Measured specific heat over $T$ is plotted across a wide $T$ range in (**A**). The border between regions II and III is clearly signified by a maximum in $C_p/T$. Within the C-CDW region the data are fitted by the sum of contributions from the crystal lattice and the electrons. For region I the lattice contribution follows its asymptotic $T^3$ form and the power law exponent for the electronic contribution $n_C$ is straightforwardly obtained from the fit, as plotted in (**B**) for temperatures up to 10 K. In regions II and III the lattice contribution is large, but the $T$ dependence of the electronic term can still be modelled using two effective values for $n_C$, as shown in (**C**). It is notable that region II has a significantly larger $n_C$ value than both regions I and III.

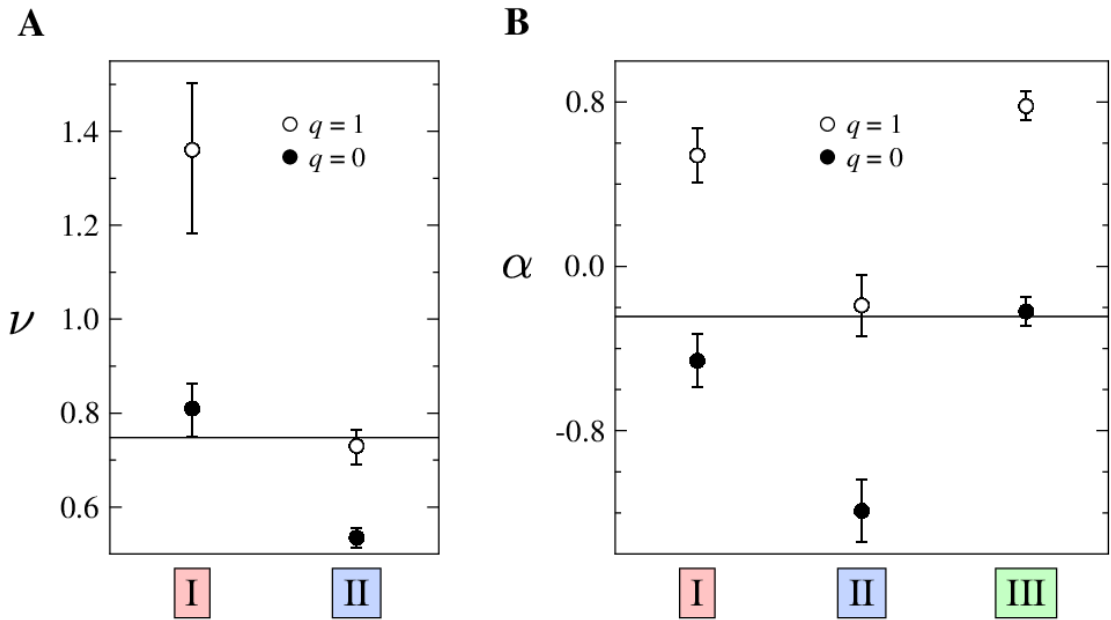

**Fig. 5. Critical parameters obtained in the three regions.** Comparison of the exponent $\nu$ from LF-µSR (**A**) and the exponent $\alpha$ from specific heat (**B**) with the O(4) values for the $Z_2$ QSL model (shown as solid lines). In each region the obtained values obtained are shown for the two cases $q = 0$ and $q = 1$. For regions I and III the measurements are best described by $q = 0$. For region II the value $q = 1$ is obtained from both measurements.

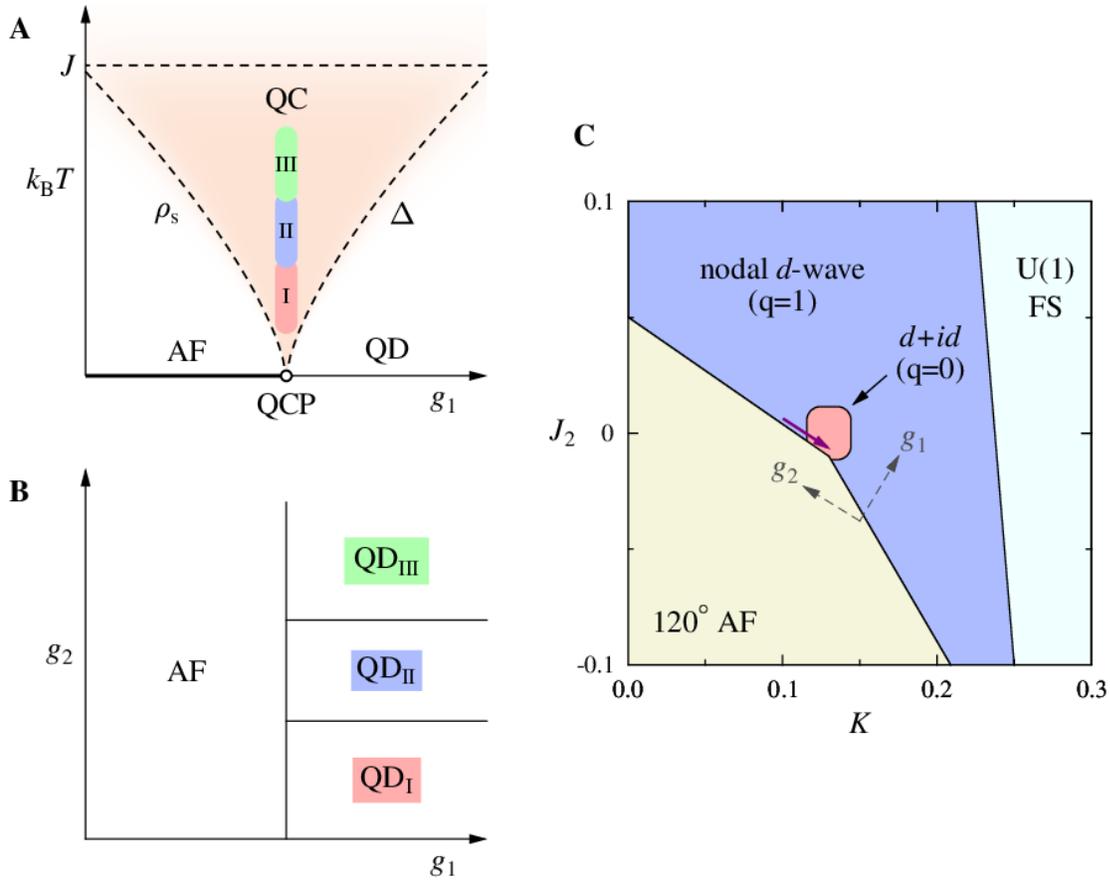

**Fig. 6. Phenomenological interpretation of the thermal phases and comparison with a reported QSL phase diagram for the triangular lattice.** (**A**) The QC spin liquid model has a QCP between an AF phase (in this case the 120° non-collinear state, shown by the thick line) and a gapped quantum disordered (QD) phase (in this case a $Z_2$ QSL). The QPT is tuned by a control parameter $g_1$ and at the QCP, the spin stiffness $\rho_s$ and the QSL gap $\Delta$ both reduce to zero. The experiments at finite temperature $T$ indicate QSL phases that are gapless and thus close to being directly above the QCP in the QC region, suggesting that the actual $g_1$ value closely matches the position of the QCP and does not have a strong $T$ dependence. The fan-shaped QC region is bounded by three crossovers (dashed lines). On the low side these reflect $\rho_s$ and $\Delta$. The exchange coupling $J$ would normally provide the high limit, but $T_{nC}$ actually occurs first here. (**B**) A second parameter $g_2$ is introduced to reflect the stabilization of different QSL phases. (**C**) A schematic representation of the relevant part of the phase diagram for the Heisenberg antiferromagnetic triangular lattice, as reported from variational studies (*40*). The control parameters are the second-nearest-neighbor exchange $J_2$ and the four-site ring exchange $K$, with both parameters normalized to the nearest-neighbor exchange $J$. Within a low $J_2$ region close to the AF state, the $d + id$ state and the nodal $d$-wave state are very close in energy, leading to a pocket of stability for the $d + id$ state. Dashed lines indicate a proposed mapping of the phenomenological control parameters $g_1$ and $g_2$ to the $K$-$J_2$ parameter space of this model. The purple arrow shows a possible path for the parameters of 1T-TaS$_2$ when cooling through regions II and I that would be consistent with our results.

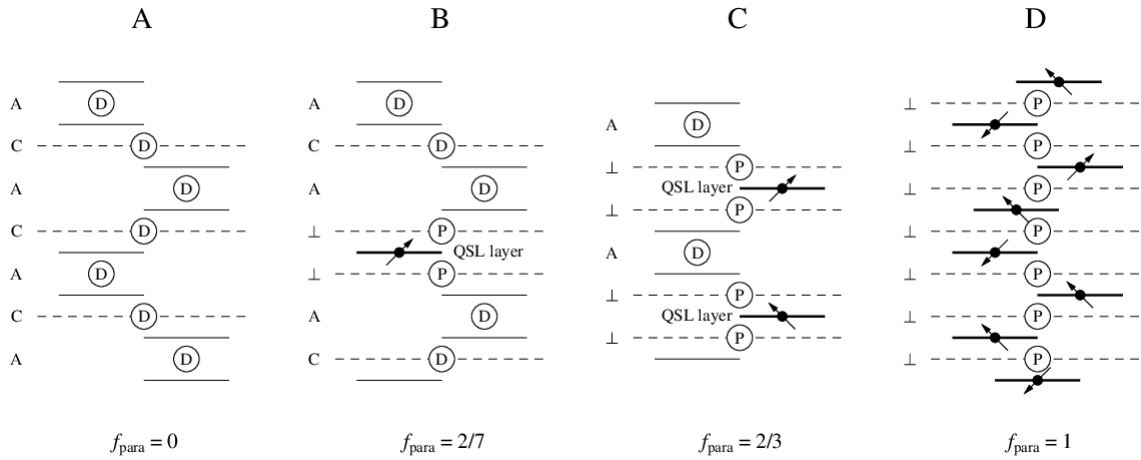

**Fig. 7. Layer stacking and the muon relaxation signal.** (**A**) Regular ACACAC stacking of the triangular-lattice layers of Star-of-David (SD) clusters produces completely spin-paired bilayers that do not support the QSL state. Cleavage at C planes (dashed lines) in such a structure results in bilayer surfaces. Muon sites in this case all see diamagnetic environments (shown as D here). (**B**) A defect in the SD stacking sequence at the level of one in seven layers on average gives a sequence such as ACA⊥⊥AC for which 1 in 4 cleaved surfaces are unpaired monolayers that can support the QSL state. This ratio matches that seen in a recent STM study (24). The muon stopping sites in this case split between diamagnetic (D) and paramagnetic (P) environments, with $f_{para}$, the P to D site ratio, being 2/7 for this concentration of stacking defects. (**C**) and (**D**) show example sequences with greater concentrations of the electronic stacking defects giving more unpaired QSL layers and larger values of $f_{para}$.

# Supplementary Materials for

## Quantum Phases and Spin Liquid Properties of 1T-TaS$_2$

Samuel Mañas-Valero, Benjamin M. Huddart, Tom Lancaster, Eugenio Coronado, Francis L. Pratt*

*Corresponding author. Email: francis.pratt@stfc.ac.uk

**This PDF file includes:**

Supplementary Text for Materials and Methods
Figs. S1 to S6
Tables S1 to S4

# Supplementary Text

## Materials and Methods
1. Sample Preparation and Characterization

### *Crystal growth*
High-quality 1T-TaS$_2$ bulk crystals (Fig.S1a,b) were synthesized by chemical vapour transport and characterized using elemental analysis, ICP mass spectrometry and XRD.

The crystal growth conditions were as follows: first, powders of Ta (99.99 %, Alfa-Aesar) and sulfur (99.8 %, Sigma-Aldrich) were mixed in a stoichiometric ratio, sealed in an evacuated quartz ampoule (length: 250 cm, internal diameter: 14 mm; P: $2 \cdot 10^{-5}$ mbar) and heated from room temperature to 910 °C in 3 hours. The temperature was then kept constant for ten days and finally allowed to cool down naturally. In order to grow larger crystals, one gram of the obtained material was mixed with iodine (99.999 %, Sigma-Aldrich; [I$_2$] = 1 mg/cm$^3$), and sealed in an evacuated quartz ampoule (length: 500 mm, internal diameter: 14 mm; P: $2 \cdot 10^{-5}$ mbar). The quartz tube was placed inside a three-zone furnace with the material in the leftmost zone. The other two zones were heated up in 3 hours from room temperature to 950 °C and kept at this temperature for two days. After this, the leftmost side was heated up to 925 °C in 3 hours and there was established in the three-zone furnace a gradient of 950 °C/900 °C/925 °C. This temperature profile was kept constant for 20 days and the final stage was a rapid quench into water.

The weight concentration of elements obtained by inductively coupled plasma optical emission spectroscopy (ICP-OES) was Ta: (74.6 ± 1.4) % and S: (25.6 ± 0.5) %. These values are in good agreement with the expected ones (Ta: 73.8 % and S: 26.1 %). Phase purity was confirmed by the refinement of the X-ray pattern (measured with a PANalytical Empyrean X-ray platform with Cu radiation source; see Section 5) to the previously reported structure of 1T-TaS$_2$ (ICSD 85323) using the X´Pert Highscore Plus program. The hexagonal crystal system with a P-3m1 space group was obtained and the unit cell was determined to be α = β = 90°, γ = 120°, a = b = 3.3662(4) Å and c = 5.8975(7) Å. The obtained results are in accordance with the ones reported in the literature (*49*).

### *Magnetic susceptibility*
Magnetic susceptibility was measured using a Quantum Design MPMS SQUID system. The measured susceptibility is shown in Fig. S1c. As a basic characterisation to allow comparison with previous reports, the data below $T_{nC}$ were fitted to the form $\chi = \chi_0 + C/(T+\theta)$. Parameters for the sample in this study are compared with those of some previous studies in Table S1. It can be seen that the sample used in the present study shows the largest Pauli-like term and the smallest impurity term, corresponding to a weakly interacting spin ½ concentration of 0.02 %. A close inspection of Fig. S1c shows an additional weak feature around 50 K where there is a small drop in χ. There are no comparable features in any of our other measurements and we assign this drop to a modification of the impurity term. This may be related to a change in the magnetic environment of the impurity, e.g. we can speculate that it reflects an impurity in a host diamagnetic layer that becomes a paramagnetic layer above 50 K.

*Electronic transport measurements*

In order to confirm the hysteretic behavior related to the commensuration of the CDW and compare it with previous results in the literature, large crystals of 1T-TaS$_2$ were contacted with silver conductive paste (from RS, Stock No. 186-3593) in an in-plane (*ab* plane) 4 probe configuration using platinum wires. The transport measurements were carried out in a Quantum Design PPMS-9 with a cooling/warming rate of 1K/min. The resistivity values as well as the thermal dependence (Fig. S1d) are in agreement with the previous results in the literature (*18*, *51*), showing the formation of the commensurate charge density wave superlattice at around 200 K.

*Muon measurements*

The μSR experiments were carried out on the HiFi instrument at the ISIS Neutron and Muon Source. A mosaic sample was made from several large crystals of 1T-TaS$_2$ (0.7272 grams in total) all oriented with the *ab* plane parallel to the surface (Fig. S1b). For the measurements the mosaic was mounted on a silver sample plate and covered with a thin silver foil. This was placed in a closed cycle refrigerator and a beam of positive surface muons was implanted into the sample. Data analysis was carried out using the WiMDA program (*49*).

The muon probe is fully spin polarised on implantation and the forward/backward asymmetry of the detected muon decay positrons $a(t)$, reflects the time dependent polarisation of the muon spin. In the ZF measurements (Fig. S1e) the relaxing component of $a(t)$ was fitted to the product of a Gaussian and a Lorentzian term. The Gaussian term reflects nuclear dipolar relaxation contributions and was estimated at high $T$ and then kept fixed as $T$ varied. The relaxation rate of the Lorentzian term $\lambda$ reflects the electronic contribution to the muon spin relaxation that varies significantly with $T$. The electronic contribution is in a fast fluctuation regime where $\lambda$ is proportional to the electronic spin fluctuation time $\tau$. In the LF measurements the magnetic field was applied perpendicular to the *ab* plane of the crystals.

Detailed measurements of the field and temperature dependence of the muon spin relaxation were made in this study, counting $5 \times 10^7$ muon decay events in each μSR spectrum to get sufficient accuracy in the fitted parameters to estimate critical exponents that can be compared with those of QSL models.

*Specific heat*

Specific heat was measured using a Quantum Design PPMS. The results obtained over a wide temperature range are shown in Fig. 4A. For the data analysis a good representation of the lattice contribution is required. For the low temperature range within region I a simple Debye $T^3$ dependence is appropriate (Fig. 4B). At higher temperatures the data become affected by the anisotropy of the acoustic phonons and the excitation of optical phonons. In regions II and III the phonon term is modelled by the sum of an anisotropic Debye term (*52*) representing the acoustic phonons and two Einstein terms to represent optical phonons

$$C_p^{\text{lat}}/R = 3\, F_D(\theta_D, r) + 4\, F_E(\theta_{E,1}) + 2\, F_E(\theta_{E,2})$$

where $F_D$ is the anisotropic version of the Debye function (*53*) for anisotropy $r$ and $F_E$ is a standard Einstein specific heat function. The four effective phonon parameters $\theta_D$, $r$, $\theta_{E,1}$ and $\theta_{E,2}$ were determined by fitting the data in region II between 30 K and 110 K, taking the spinon term to follow $n_C = 2.19$, the appropriate power law deduced from the spinon diffusion analysis in region II. This procedure works well in describing the data (Fig. 4A) and allows the clear drop in $n_C$ on going from region II to region III to be quantified (Fig. 4C). Parameters obtained for the phonon terms are listed in Table S4, together with an overall comparison of the experimental parameters obtained in this study.

2. LF-μSR and Spin Diffusion

When LF is applied the nuclear contribution to the relaxation is fully quenched at fields above a few mT leaving just the electronic contribution. At higher fields a crossover to the slow fluctuation regime can take place. The relaxation rate $\lambda(B_{LF})$ has a field dependence reflecting the spectral density function of the relaxation process $S(\omega)$, the spectral density being derived from the Fourier transform of the spin autocorrelation function $P(t)$, i.e.

$$\lambda(B_{LF}) \propto S(\omega) = \int_0^\infty P(t) \cos \omega t \; dt \tag{4}$$

where $\omega$ is the probe frequency that scales with $B_{LF}$. To fit $\lambda(B_{LF})$ in this case a model is considered in which spinons diffuse on an $n$-dimensional lattice where $n=1$ is a chain, $n=2$ is a square lattice and $n=3$ is a cubic lattice. The spin autocorrelation function is then given by (28)

$$P_n(t) = [e^{-2tD_{nD}} I_0(2tD_{nD})]^n \tag{5}$$

where $I_0$ is the zeroth order Bessel function and the diffusion rate is $D_{nD}$ and the corresponding spectral density for $n$ dimensional diffusion is

$$S_{\text{diff}}(\omega, n) = \int_0^\infty P_n(t) \cos \omega t \; dt \tag{6}$$

Comparison was made between the measured data and the $\lambda(B_{LF})$ form predicted by the 1D, 2D and 3D spinon diffusion models (Fig. 2). The 1D model leads to a $B_{LF}^{-1/2}$ form for $\lambda(B_{LF})$ that is clearly inconsistent with the data, whereas 2D and 3D models are much closer to the measurements. Careful comparison with the data shows that at every temperature the 2D model gives a significantly better fit than the 3D model. Thus $D_{2D}$ versus temperature may be obtained from LF scans at a series of temperatures.

In general, both diffusive and localised excitations contribute to $\lambda(B_{LF})$. Localised excitations are represented by a Lorentzian spectral density centred on zero field, whereas fast diffusing excitations have spectral density $S_{\text{diff}}(\omega,n)$ extending to higher fields. When the isotropic hyperfine coupling $\bar{A}$ dominates over the dipolar coupling, the diffusive contribution to the LF relaxation rate from the spin ½ electronic spins can be expressed as

$$\lambda_{\text{diff}}(B_{LF}) = f_0 \, \bar{A}^2/2 \; S_{\text{diff}}(\omega,2) \tag{7}$$

where the probe frequency is the electronic Larmor frequency $\omega = \gamma_e B_{LF}$, with $\gamma_e/2\pi = 2.802\times 10^{10}$ s$^{-1}$T$^{-1}$. In (7) $f_0$ represents the product $f_{\text{diff}} f_{\text{para}}$, where $f_{\text{diff}}$ is a factor ranging from 0 to 1 to describe the diffusive fraction of the overall spectral density and $f_{\text{para}}$ is the fraction of muon sites (ranging from 0 to 1) that sense TaS$_2$ layers with unpaired spins. Both the diffusion rate $D_{2D}$ and $f_0$ are evaluated via (7) at each temperature, as shown in Fig.3A,B. The diffusion rate depends only on the shape of $S_{\text{diff}}(\omega,2)$, whereas obtaining an absolute value for $f_0$ depends on estimating the value of $\bar{A}$, which is discussed in the following section.

## 3. Muon Sites and Hyperfine Coupling

In order to identify the muon sites and their hyperfine coupling to the unpaired spin, calculations were made with plane-wave density functional theory (DFT) using the CASTEP program (*55*) with the generalised gradient approximation and the PBE functional (*56*). Calculations were carried out on a $\sqrt{13} \times \sqrt{13} \times 2$ supercell comprising two C-CDW unit cells stacked along the *c*-axis (the doubling of the *c*-axis is necessary to ensure sufficient isolation of the implanted muon from its periodic images). We use a plane-wave basis set cut off energy of 500 eV and a $5 \times 5 \times 5$ Monkhurst-Pack (*57*) grid for *k*-point sampling, resulting in total energies that converge to an accuracy of 0.01 eV per cell. Muons, represented by a hydrogen pseudopotential, were initialised in 59 random positions within this supercell, generated by requiring initial positions being at least 0.25 Å away each other and from any of the atoms in the cell. Neutral muonium was used to provide an unpaired spin within the structure and the hyperfine coupling was evaluated for each relaxed site configuration (calculations for $\mu^+$ were found to give negligible hyperfine coupling). In order to be compatible with the expected Mott-Hubbard insulator electronic state of a QSL for calculations of the hyperfine coupling of the muon to the QSL layers, the system was constrained to be an insulator. Calculations where the whole system was allowed to adopt a metallic ground state, an approach taken in several other DFT calculations on bulk 1*T*-TaS$_2$ (*52*, *58*, *59*), yielded negligible spin density on all atoms and negligible hyperfine coupling at the muon site, which is clearly not compatible with experiment.

The obtained final sites form four distinct groups (Fig. S2). Muons in site 1 (Fig. S2a) bond to one S atom in each of two adjacent TaS$_2$ layers forming a linear S-μ-S state with unequal μ-S bond lengths of 1.56 Å and 1.65 Å. Site 1 comprises two crystallographically distinct subclasses of sites that are made distinct by the CDW distortions. However, the local geometry of the muon is nearly identical in the two cases, resulting in similar hyperfine coupling. In site 2 (Fig. S2b) the muon is bonded to only a single S atom (with a μ-S bond length of 1.4 Å), with these sites being around 0.03 eV higher in energy than site 1. In site 3 (Fig. S2c), the muon stops inside the S layer, slightly displaced from the centre of the triangle defined by three S atoms. For site 4 (Fig. S2d) the muon site is displaced by around 0.7 Å along the c-axis above the centre of the triangle defined by three Ta atoms.

The dipolar coupling is much smaller than the contact hyperfine coupling in each case. The hyperfine coupling is averaged over the members of each site group to give a value for each site (Table S2). The calculated hyperfine coupling constants were found to be highly sensitive to the *k*-point grid used (varying by about 20%) and the $5 \times 5 \times 5$ grid used in our calculations is insufficient to ensure full convergence of the hyperfine coupling. However, larger *k*-point grids are computationally prohibitive and this limits the accuracy of our estimates at each site to around 1 MHz. In Table S2 we also report the zero-point energy (ZPE) of the muon, determined via the harmonic approximation from the three highest frequency phonon vibrations in the lowest energy structure within each classes of stopping site, calculated using the finite displacement method.

To investigate the possibility of muon diffusion within the temperature range of the experiment, we carried out transition state searches (*60*) between pairs of distinct muon stopping sites associated with a single TaS$_2$ layer. Here, the lowest energy site within each cluster is chosen as the representative member of this cluster (Table S3). Energy barriers denoted as $i \rightarrow j$ correspond to those between adjacent sites of type *i* and *j* within the structure. Energy barriers are illustrated in Fig. S3. We find that the energy barrier between sites 1 and 2 is very close to the difference in energy between the two. This, coupled with the large ZPEs (0.35 eV and 0.51 eV respectively) of both sites means that the muon is likely to be delocalised between these two geometries, rather than them representing two distinct stopping sites. Similarly, the barrier between sites 2 and 3 is less than their ZPE. Thus sites 1, 2 and 3 are expected to form a single quantum delocalised state with average coupling 7.2 MHz (penultimate row in Table S2). For site 4 the ZPE is larger than the barrier to site 3 and thus a transition to site 3 is expected. Barriers between adjacent sites of the same type in the ab plane are all > 1 eV and hence large compared with the ZPE and thermal energy available to the muon. Long range muon diffusion is therefore expected to be strongly suppressed in this structure and thus cannot be responsible for the rapid fall in $D_{2D}$ and the $f_{\text{diff}} f_{\text{para}}$ product with temperature that is found in region III.

Since the barrier between type 1 sites of adjacent $TaS_2$ layers is very weak, we need to allow for the possibility of a further level of delocalisation across the interlayer region, linking the sites 1-3 above and 1′-3′ below the midpoint between the layers (Fig. S3b). For the case where only the upper $TaS_2$ layer has an unpaired spin, the hyperfine coupling to the spin in the upper layer is expected to be zero when the muon is at sites 2′ and 3′ (Fig. S3b). The estimated average coupling for the delocalised muon now reduces to $\bar{A} = 4.5(6)$ MHz (final row in Table S2).

4. The diminished muon response in region III

The rapid fall-off in the muon relaxation signal with increasing temperature above 110 K is not related to muon diffusion, as was discussed in section 3. The sensitivity of the muon probe state to spins in two adjacent layers was highlighted in section 3 and this characteristic is likely to be behind the reduced response in the higher temperature phase. In particular, if unpaired spins are present in two adjacent layers that show antiferromagnetic correlation, then the hyperfine couplings of the muon to the spins in the two layers will tend to cancel, leading to a loss of muon relaxation signal (Fig. S4).

5. Summary of the properties of the three thermal regions below $T_{nC}$.

Table S4 summarizes the properties of the three thermal regions.

6. Estimation of structural stacking faults.

In a layered system, the lattice planes are stacked periodically, except for stacking faults between two successive planes. In a hexagonal closed packed (HCP) structure, deformation faults and growth faults are possible on the 001 planes (i.e., in the c-axis, the direction where the planes are stacked). If α is the probability of a deformation fault and β is the formation of a growth fault, it can be proved (61, 62) that for the reflections of the type $H - K = 3N \pm 1$, where HKL are the Miller′s indexes, the following relations are given:

$$\text{L even:} \ B_{2\theta} = \frac{360}{\pi^2} \tan(\theta) \ |L| \left(\frac{d}{c}\right)^2 (3\alpha + 3\beta)$$
$$\text{L odd:} \ B_{2\theta} = \frac{360}{\pi^2} \tan(\theta) \ |L| \left(\frac{d}{c}\right)^2 (3\alpha + \beta)$$
(8)

where $B_{2\theta}$ is the full width at half maximum intensity expressed in degrees $2\theta$ on the powder pattern, HKL are the Miller′s indexes, d is the HKL spacing, $c = 2d_{002}$.

In total, the X-ray powder pattern of 6 different crystals of 1T-TaS$_2$ (Fig. S5), analysed as described in section 1, were fitted as described in equation (8). The experimental values of α and β are (0.71 ± 0.05) % and (0.27 ± 0.07) %, respectively (see Fig. S6). As well, the upper limit of stacking faults has been estimated (*i.e.*, the values of α and β when β and α, respectively, are set to zero), giving a maximum value of *ca.* 2 % (Fig. S6).

The amount of structural stacking faults (maximum 2 %) is much lower than the estimated amount of *electronic* stacking faults. Therefore, we can discard the observation of QSL layers as being a consequence of a defective material, suggesting its origin may instead arise from an irregular *electronic stacking* of the SD layers.

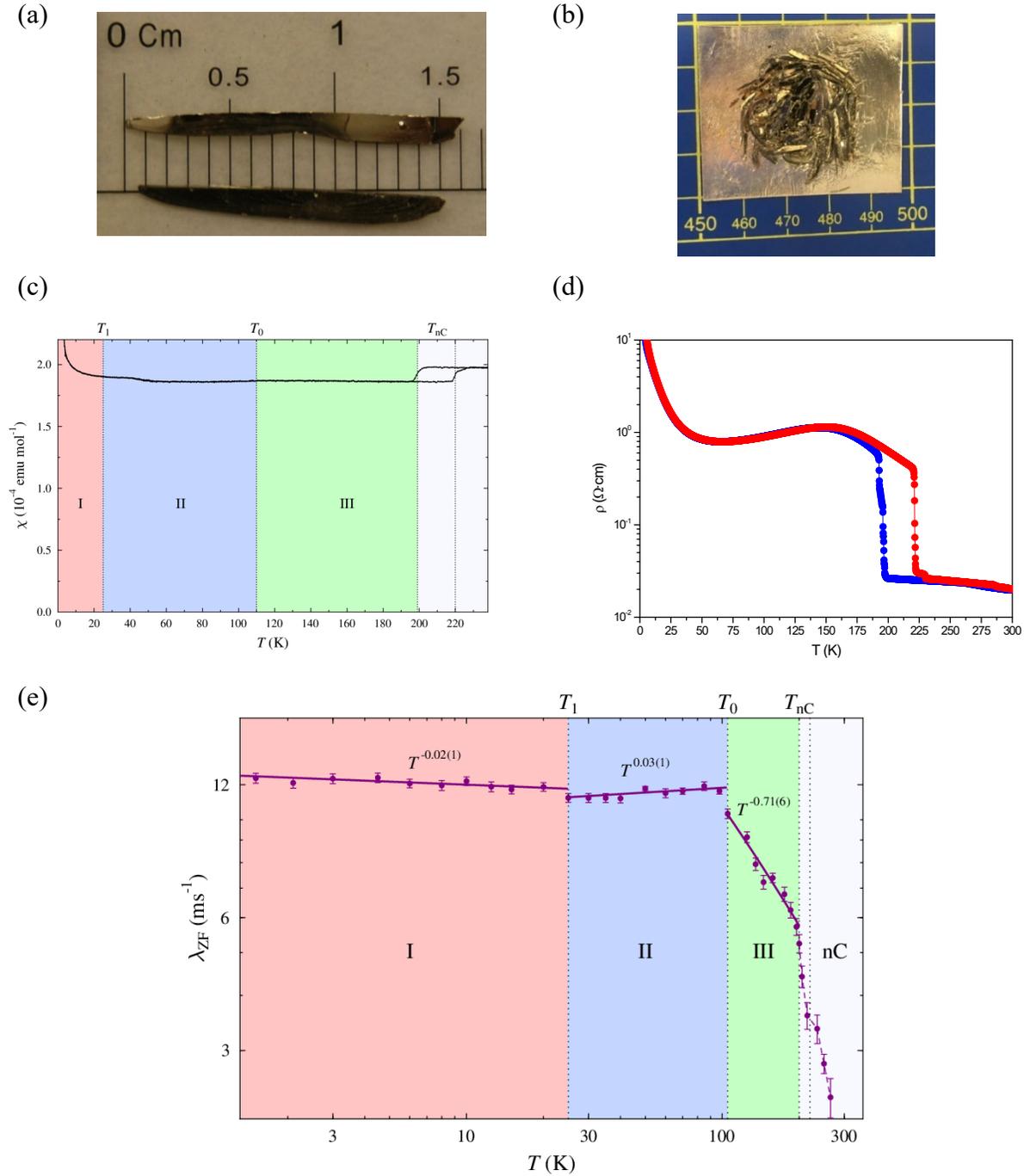

**Fig. S1.** (a) Large single crystals of 1T-TaS$_2$. (b) The mosaic sample of 1T-TaS2 crystals used in the μSR measurements. The scale is in millimetres. (c) Temperature dependent magnetic susceptibility measured under 1kOe applied magnetic field. (d) Resistivity of 1T-TaS$_2$ measured using an in-plane four-probe configuration. The cooling (blue) and warming (red) transport curves exhibit the typical hysteretic behavior around 200 K due to the formation of a commensurate charge density wave. (e) The measured T dependent muon spin relaxation rate in zero field, showing significant change in behaviour at T$_0$ and a more subtle change at T$_1$.

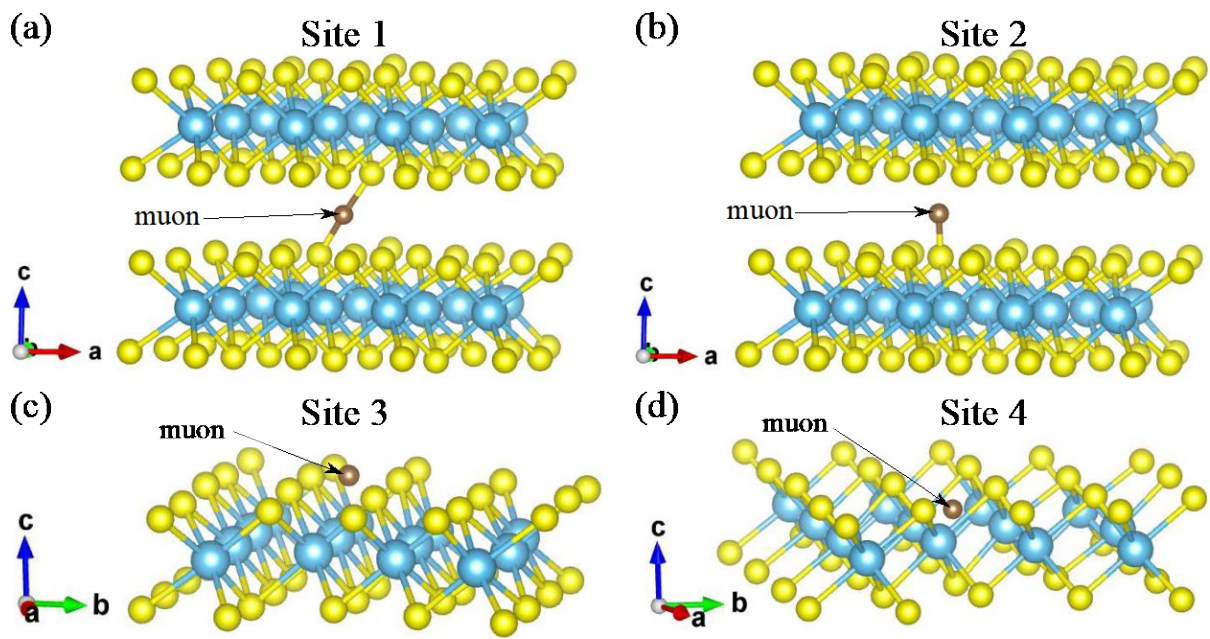

**Fig. S2.** Muon stopping sites calculated using density functional theory. Ta and S are represented by blue and yellow spheres respectively. Sites 1 and 2 occupy the interlayer region. Site 3 is associated with the S layer and site 4 with the Ta layer.

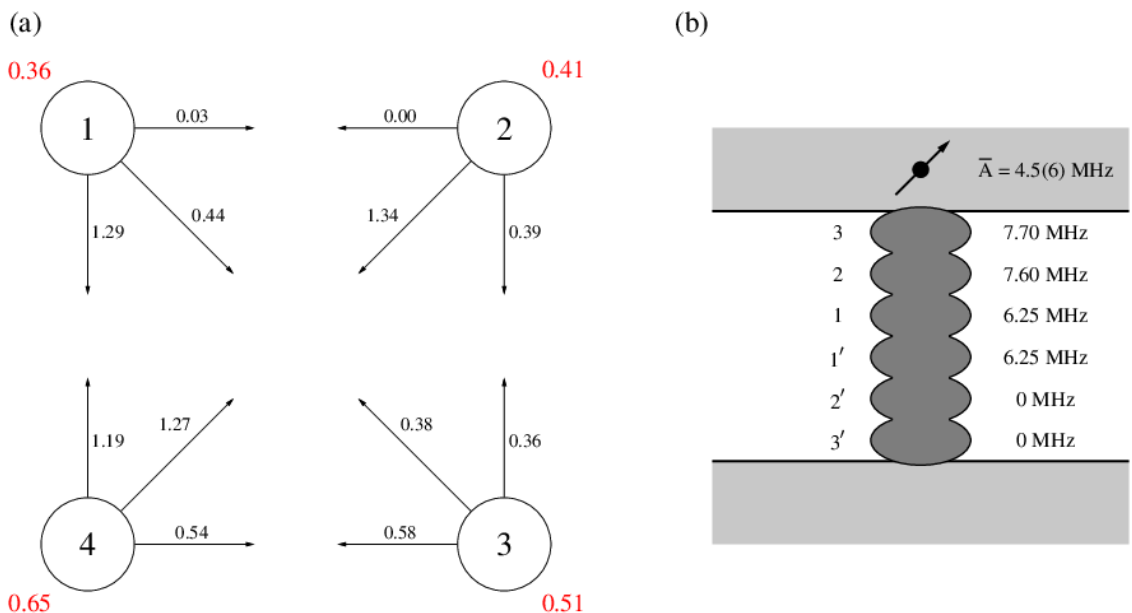

**Fig. S3.** (a) Effective barrier heights in eV between the sites, taking the site energy differences into account. The corresponding ZPE of each site is shown in red. It can be seen that site 4 will be unstable against a transition to site 3 and that sites 1, 2 and 3 will combine to form a single quantum delocalised site, whose average coupling is listed in the penultimate row of Table S2. (b) Sites 1 and 1′ are spatially very close with a very small barrier between them, so a further level of muon delocalisation will take place across the interlayer region combining sites 1-3 and 1′ to 3′. In calculating the resultant hyperfine coupling, it is assumed that the central sites 1 and 1′ couple equally well to either layer, whereas sites 2 and 3 and 2′ and 3′ will only couple to the adjacent layer. Taking this into account, the average value for hyperfine coupling of the extended muon state to an unpaired spin on a single layer is estimated to be $\bar{A}$ = 4.5(6) MHz.

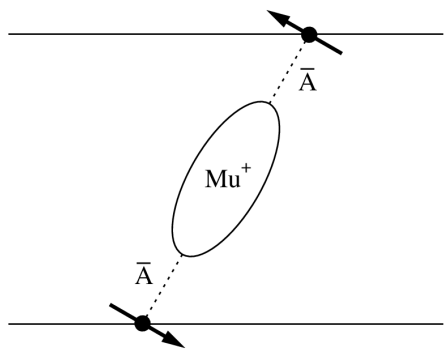

**Fig. S4.** The extended muon probe state coupling to two unpaired electronic spins in two adjacent layers that show AF spin correlation. The hyperfine couplings of the two electronic spins to the muon will cancel out in this case.

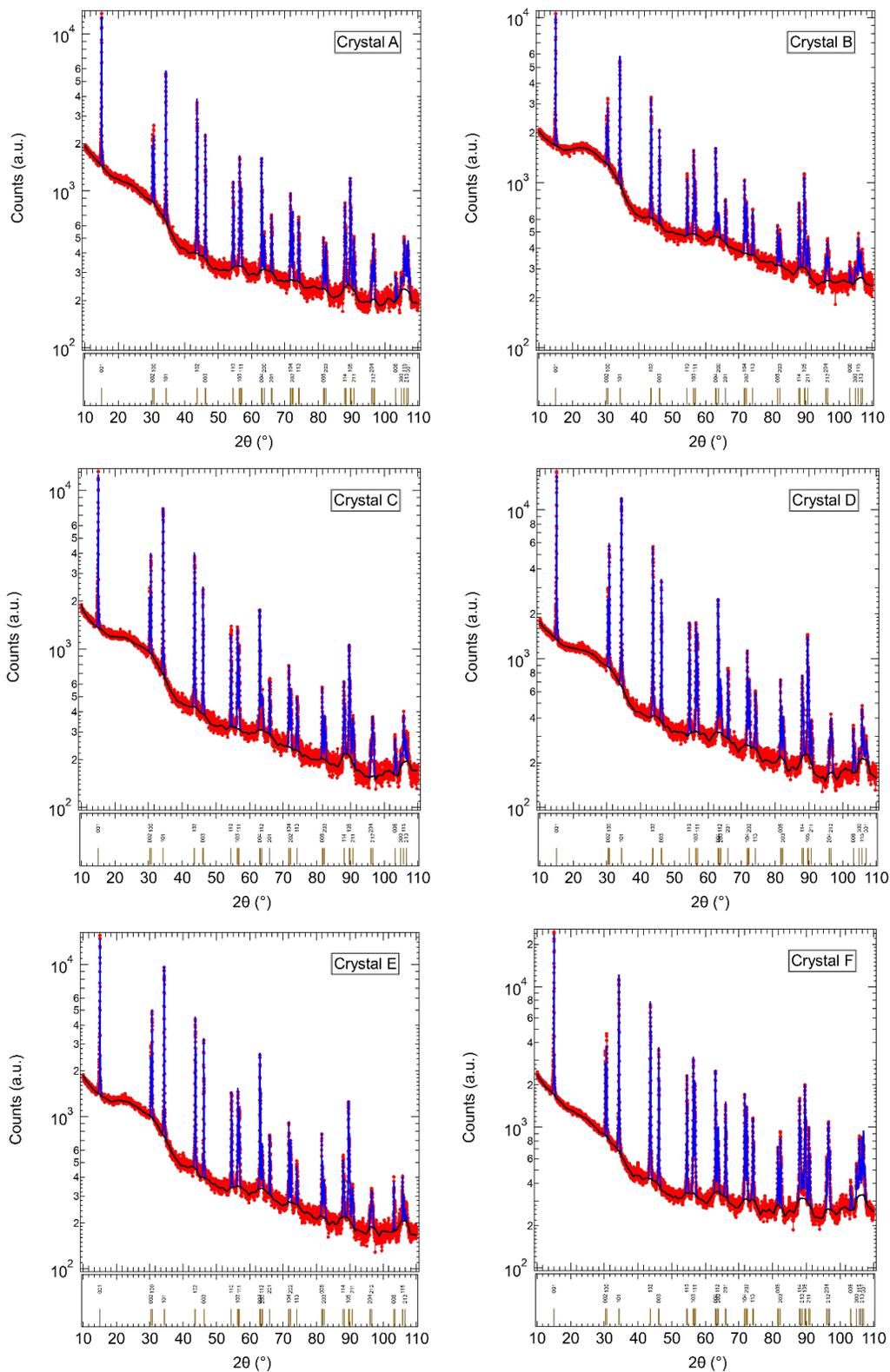

**Fig. S5.** Experimental X-ray powder diffraction pattern (red) and corresponding fit (peaks in blue and background in black) for the different 1T-TaS$_2$ crystals ground into powder, as employed for determining the probability of stacking faults.

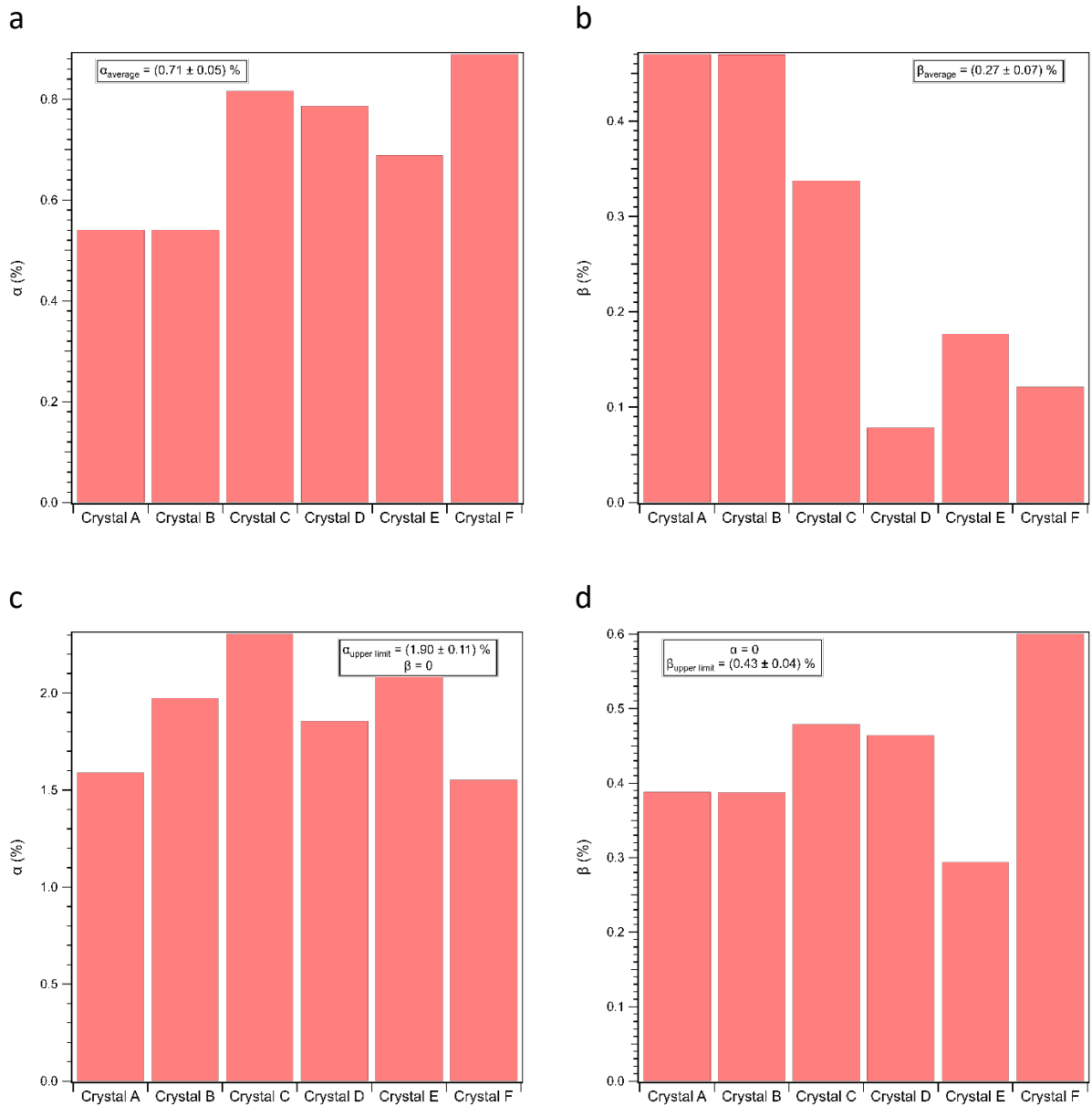

**Fig. S6.** Estimation of the probability of stacking faults for six different crystals of 1T-TaS$_2$: (a) α deformation fault, (b) β growth fault. The upper limit for probability of stacking faults: (c) α deformation fault and (d) β growth fault.

|  | $\Delta\chi$ at $T_{nC}$ ($10^{-4}$ emu mol$^{-1}$) | $\chi_0$ ($10^{-4}$ emu mol$^{-1}$) | $C$ ($10^{-4}$ emu mol$^{-1}$ K$^{-1}$) | $n$ (%) | $\theta$ (K) |
|---|---|---|---|---|---|
| DiSalvo (*54*) | 0.12 | −0.5 | 2.0 | 0.05 | −0.4 |
| Kratochvilova (*18*) | ∼ 0.2 | −0.7 to +0.7 | 6 to 17 | 0.15 to 0.5 | 0.02 |
| Klanjšek (*19*) | ∼ 0.15 | 0.3 | 2.3 | 0.06 |  |
| Ribak (*20*) | 0.18 | 0.4 | 1.5 | 0.04 | 2.1 |
| Present study | 0.12 | 1.9 | 1.0 | 0.02 | 0.6 |

**Table S1.** Comparison of the magnetic parameters obtained in the present study with some previously reported values. $\Delta\chi$ is the drop in susceptibility at the transition to the commensurate CDW state. $\chi_0$ is the magnitude of the temperature-independent Pauli-like term and C is the magnitude of the Curie term, with *n* showing the equivalent concentration of impurities for $S = 1/2$ and $g = 2$. The final column shows $\theta$, the Weiss constant associated with the Curie term.

| Site Type/State | Energy (eV) | Fraction (%) | Average hyperfine coupling (MHz) | Zero-point energy (eV) |
| --- | --- | --- | --- | --- |
| 1 | 0.03 | 34 | 6.25 | 0.36 |
| 2 | 0.06 | 12 | 7.60 | 0.41 |
| 3 | 0.09 | 47 | 7.70 | 0.51 |
| 4 | 0.13 | 7 | -0.92 | 0.65 |
| Sites 1-3 (average) | | | 7.16 | |
| Full interlayer state | | | 4.5(6) | |

**Table S2.** Muon sites and their hyperfine couplings to a single layer, plus zero-point energies.

| Transition | Energy barrier (eV) |
|---|---|
| 1 → 1 | 1.14 |
| 1 → 2 | 0.03 |
| 1 → 3 | 0.44 |
| 1 → 4 | 1.29 |
| 2 → 2 | 1.25 |
| 2 → 3 | 0.39 |
| 2 → 4 | 1.34 |
| 3 → 3 | 1.49 |
| 3 → 4 | 0.58 |
| 4 → 4 | 2.01 |

**Table S3.** Energy barriers between muon stopping sites.

| Region | I | II | III |
|---|---|---|---|
| $\theta_D$ (K) | 150.7(6) | 145(3) | 145(3) |
| $r$ |  | 2.1(3) | 2.1(3) |
| $\theta_{E,1}$ (K) |  | 518(16) | 518(16) |
| $\theta_{E,2}$ (K) |  | 313(3) | 313(3) |
| $T_{max}$ (K) | 25 | 110 | 200 |
| $c_{el}(T_{max}) / R$ | 0.019(3) | 0.49(9) | 1.0(2) |
| $n_C$ | 1.46(13) | 2.19(15) | 1.22(7) |
| $n_D$ | 0.47(17) | 1.74(14) | -4(1) |
| $\alpha$ | -0.46(13) | -0.19(15) (from $\nu$) | -0.22(7) |
| $\nu$ | 0.82(8) | 0.72(6) | 0.74(2) (from $\alpha$) |
| $q$ | 0 | 1 | 0 |

**Table S4.** Summary of the properties of the three regions below $T_{nC}$ obtained from this study.